# Twist-bend nematic liquid crystals in high magnetic fields

P. K. Challa[1], V. Borshch[2], O. Parri[3], C. T. Imrie[4] S.N. Sprunt[1], J.T. Gleeson[1], O.D. Lavrentovich[2], and A. Jákli[2]

[1] Department of Physics, Kent State University, Kent, OH 44242, USA

[2] Liquid Crystal Institute and Chemical Physics Interdisciplinary Program, Kent State University, Kent, OH 44242, USA`

[3] Merck Chemicals Ltd., Chilworth Technical Centre, University Parkway, Southampton SO16 7QD, UK

[4] Department of Chemistry, School of Natural and Computing Sciences, University of Aberdeen, AB24 3UE, Scotland

## Abstract:

*We present magneto-optic measurements on two materials that form the recently discovered twist-bend nematic ($N_{tb}$) phase. This intriguing state of matter represents a new fluid phase that is orientationally anisotropic in three directions and also exhibits translational order with periodicity several times larger than the molecular size. $N_{TB}$ materials may also spontaneously form a visible, macroscopic stripe texture. We show that the optical stripe texture can be persistently inhibited by a magnetic field, and a 25T external magnetic field depresses the $N$-$N_{tb}$ phase transition temperature by almost 1ºC. We propose a quantitative mechanism to account for this shift and suggest a Helfrich-Hurault-type mechanism for the optical stripe formation.*

A nematic (N) liquid crystalline (LC) phase of achiral rod-shaped molecules is characterized by a long-range orientational and short-range positional order where the direction of the long axis of single molecules is parallel to the optical axis of the medium. [1] An existence of a "twist-bend" (TB) nematic phase in which the molecular director exhibits periodic twist and bend deformations following the line of an oblique helicoid, with the optical axis being the axis of the helicoid, was also proposed long time ago by Meyer [2] and in subsequent theoretical and simulation works [3], [4], [5], [6]. Recently, dimeric compounds formed by two rigid rod-like cores linked together by a flexible aliphatic chain with odd numbered carbon atoms were found to show a transition from a nematic phase to a birefringent striped phase [7] with periodicity in the micron range. This state was originally identified as a smectic A phase [8], but x-ray studies have since ruled this out [9] and the striped state was tentatively labeled as an "$N_x$" phase. [7] $^2$H nuclear magnetic resonance spectroscopy showed that the $N_x$ phase has local chiral order [10–12], although the compounds are *achiral*. $N_x$ phase recently was also found in chiral and doped dimers. [13] [14] Freedericksz transition measurements and dynamic light scattering revealed an unusually low bend elastic constant in the nematic phase just above the N-$N_x$ phase transition. [15], [16] The $N_x$ phase also exhibits linear (polar) switching under applied electric fields with a few microseconds response time [17], [18], which was attributed to a twist-bend structure with a periodicity of a few molecular lengths (~10 nm). [19] The surprisingly small length scale was indeed directly demonstrated via freeze fracture transmission electron microscopy (FFTEM) [16] [20] to be 8-9 nm. The FFTEM textures also reveal distinctive, asymmetric Bouligand arches, verifying that the nanoscale periodic structure is indeed formed by twist-bend deformations of the molecular director [16].

Although the $N_x$ phase has now been positively identified as the twist-bend nematic ($N_{tb}$) phase (observed also in rigid bent-core material [20]), important open questions remain. Among these are the origin of the macroscopic (optical) stripe pattern and the magnitude and behavior of the elastic constants of the $N_{tb}$ phase. In this Letter we present high field magneto-optic studies of two liquid crystals exhibiting the $N_{TB}$ phase. We demonstrate that a large magnetic field can suppress the N-$N_{TB}$ transition temperature by almost 1ºC, and a sufficiently large field can persistently remove the optical stripe texture. We propose a coarse-grain model to explain the optical stripes and to estimate the elastic constants of the $N_{tb}$ phase.

We studied two materials: (1) KA(0.2), which is a 6 component formulation with 20 mol% methylene linked dimer 1″, 9″-bis(4-cyano-2′-fluorobiphenyl-4′-yl)nonane (CBF9CBF) added to a base mixture composed of five odd-membered liquid crystal dimers with ether linkages containing substituted biphenyl mesogenic groups [15]; and (2) CB7CB, which is a single-component compound 1",7"-bis(4-cyanobiphenyl-4'-yl)heptane. [8] The N-$N_{tb}$ transition temperatures are 37.4ºC and 103.4ºC for KA(0.2) and CB7CB, respectively. The materials were loaded into glass sandwich cells treated for planar alignment with cell spacing $d$ = 10μm (KA(0.2)) or 5μm (CB7CB). The cells were held in a temperature controlled oven which was placed at the center of the vertical bore of the 25T split-helix resistive solenoid magnet [21] at the National High Magnetic Field Laboratory (NHMFL).

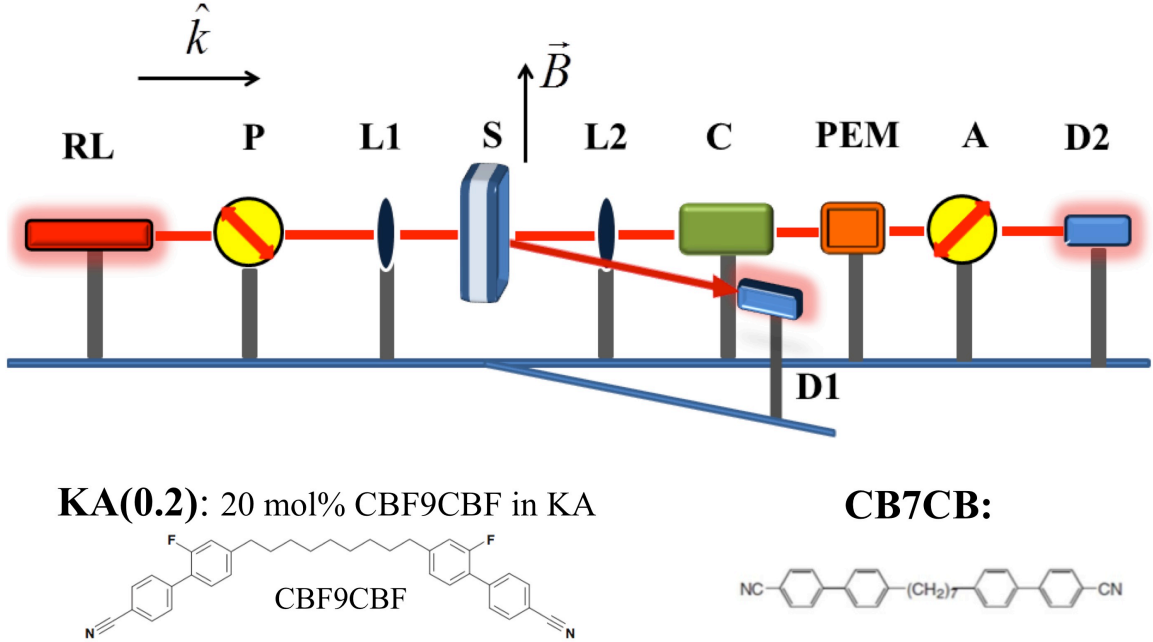

*Figure 1: Schematic of the optical setup and molecular structures of the materials KA(0.2) and CB7CB. RL: 30mW λ=632.8 nm He-Ne laser; S: LC sample placed between crossed polarizers (P and A) oriented at ±45° with respect to the vertical magnetic field; L1 and L2 are +500mm focal length lenses focusing the laser spot to ~50μm diameter. C: a compensator to correct for residual birefringence; PEM: photo-elastic modulator; D1 and D2: photodetectors positioned at the 1st order diffraction maximum and the direct beam, respectively.*

The two materials were studied using the optical arrangement depicted in Figure 1. This setup allows us to directly measure the phase difference, $\varphi = 2\pi \cdot \Delta n_{eff} \cdot d / \lambda$, where $\Delta n_{eff}$ is the effective birefringence of the sample [22,23] The optical stripe pattern, when present, acts as a diffraction grating; a detector (D1 in Fig. 1) is positioned at the first diffraction maximum to monitor the presence of these stripes. The temperature dependences of the diffracted intensity ($I_D$) at zero and 25T magnetic fields are shown in Figure 2(a) and the magnetic field dependencies at fixed temperatures are shown in Figure 2(b) and (d) for KA(0.2) and CB7CB, respectively.

As shown in Figs. 2(a), in both materials and in zero applied field, $I_D$ increases dramatically at the N-$N_{tb}$ phase transition. The optical stripes producing the diffraction are very regular in KA(0.2), and the diffraction spots are clearly defined up to the sixth order [see inset in Fig 4 (a)]. In CB7CB the stripes are more complex, and in addition to the clearly visible "macroscopic" stripes with a period comparable to the film thickness, one also observes a micron width stripes tilted with respect to the macroscopic stripes and often intercalated with arrays of focal conic domains [see Figure 4(b)]. When the samples cooled through the transition at 25T $I_D$ remains at least 100 times lower than recorded when the stripe texture is present in zero field.

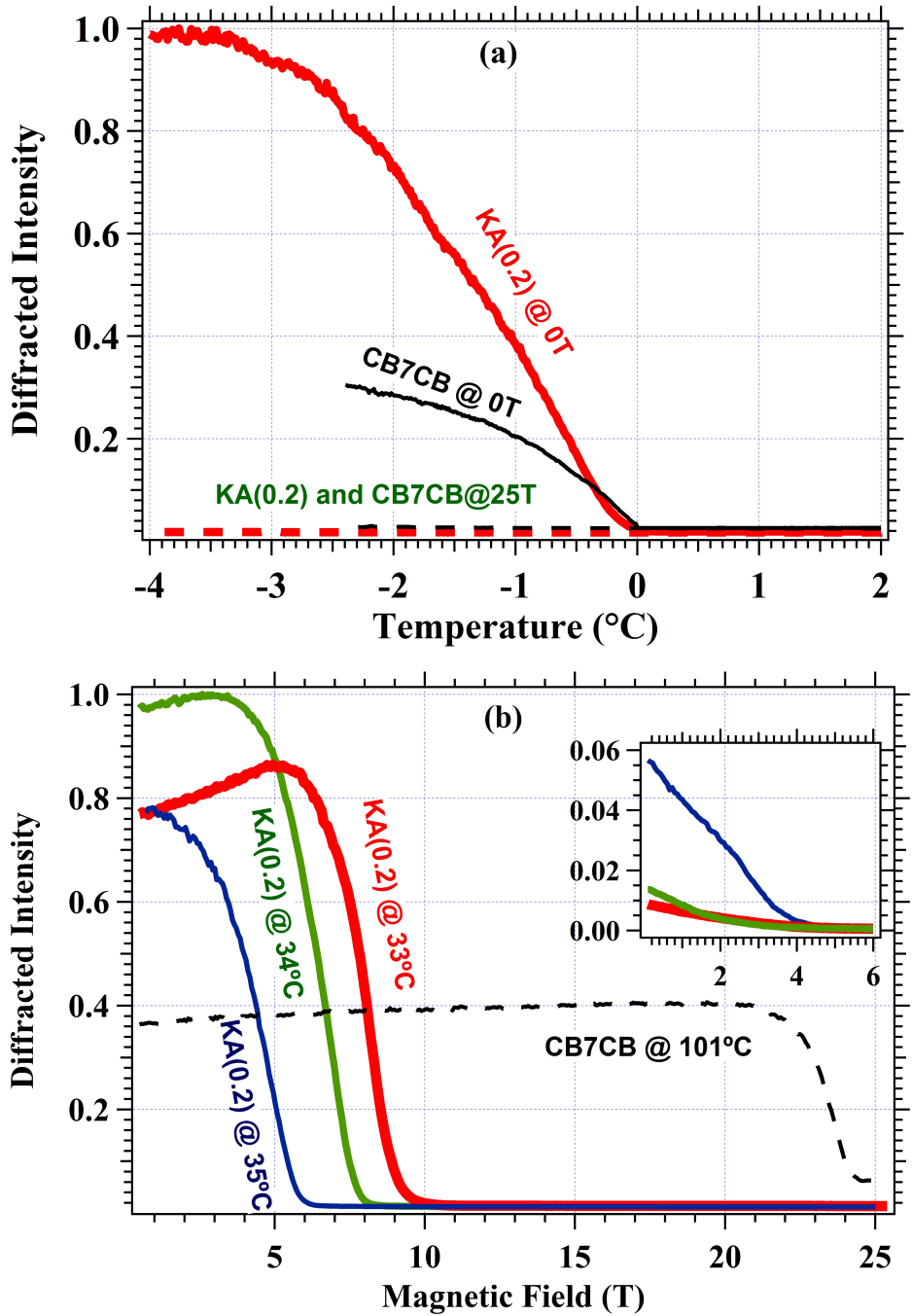

*Figure 2: Temperature and magnetic field dependence of the light intensity measured at the first diffraction peak. (a) Temperature dependences for KA(0.2) and CB7CB at 0 and 25T; (b) Magnetic field dependences for KA(0.2) at 33°C, 34°C and 35°C and for CB7CB at 101°C. Inset shows the diffracted intensity at decreasing fields for KA(0.2).*

As the field is ramped up at a rate of 5T/min at fixed temperatures below the N-N$_{tb}$ transition [Figs. 2(b)] $I_D$ dramatically decreases at a temperature-dependent threshold field. In KA(0.2) the threshold field needed to suppress the stripes increases with decreasing temperature below the (zero field) N-N$_{tb}$ transition. Interestingly at lower temperatures there is first an increase in $I_D$ before it drops to background levels. In CB7CB at 2 ºC below the transition, a much larger magnetic field was required to completely suppress the stripes. When the field is subsequently removed (at fixed temperature), the stripes only slightly appear again even after several hundred seconds have elapsed (see inset in Figure 2(b)).

The temperature and magnetic field dependences of the effective birefringence, $\Delta n_{eff}$ of both materials are non-monotonous and rather complex (see Figure 3), as they are expected to depend at least on four factors: (i) changes of the degree of orientational order; (ii) pretransitional fluctuations, (iii) formation of heliconical N$_{tb}$ structure, (iv) formation of the stripe pattern with a distorted helicoidal axis. Figure 3(a) shows the variation of $\Delta n_{eff}$ at the function of relative temperature with respect to N-N$_{tb}$ transition upon cooling at a rate of 1ºC/min. In zero field $\Delta n_{eff}$ of both materials have maxima at 3-4ºC above the N-N$_{tb}$ transition, then decrease slightly till the transition (37.4ºC for KA(0.2) and 103.4ºC for CB7CB). At the transition when the optical stripes appear, $\Delta n_{eff}$ decreases much more rapidly: without a discontinuous jump for KA(0.2), and with an abrupt drop by 0.03 for CB7CB. These indicate a second (or weak first), and a first order transition, respectively. When the sample is cooled under a 25T field, which aligns the optic axis in both phases uniformly, the true birefringence $\Delta n$ is obtained. The values of $\Delta n_{eff}$ then drop abruptly by ~0.05 at 1ºC and 0.7ºC lower than measured at zero-field for KA(0.2) and CB7CB, respectively. These show showing that the high magnetic field depresses the N-N$_{tb}$ phase transitions. Below these transitions the birefringence is only weakly temperature dependent.

Figure 3(b) shows the magnetic field dependences of $\Delta n_{eff}$ in the N$_{tb}$ phase after it was cooled at zero field below the N-N$_{tb}$ phase transition. Solid lines show the behavior for KA(0.2) at three constant temperatures, and dashed line represents the variation for CB7CB at 2.4ºC below the transition. For KA(0.2) the measured $\Delta n_{eff}$ increases strongly above 3T, 5T and 8T and saturates above 8T, 10T and 12T at 35ºC, 34ºC and 33ºC, respectively. For CB7CB at *T=101ºC*

Δ$n_{eff}$ increases slowly as the field is ramped up to about 20T, and then quite sharply between 22 and 24T. The saturated birefringence for both materials is smaller in the $N_{tb}$ than in the N phase. When the field is ramped back down to zero, the birefringence decreases only slightly and remains well above the original zero-field value (see inset to Fig 3(b) for KA(0.2)).

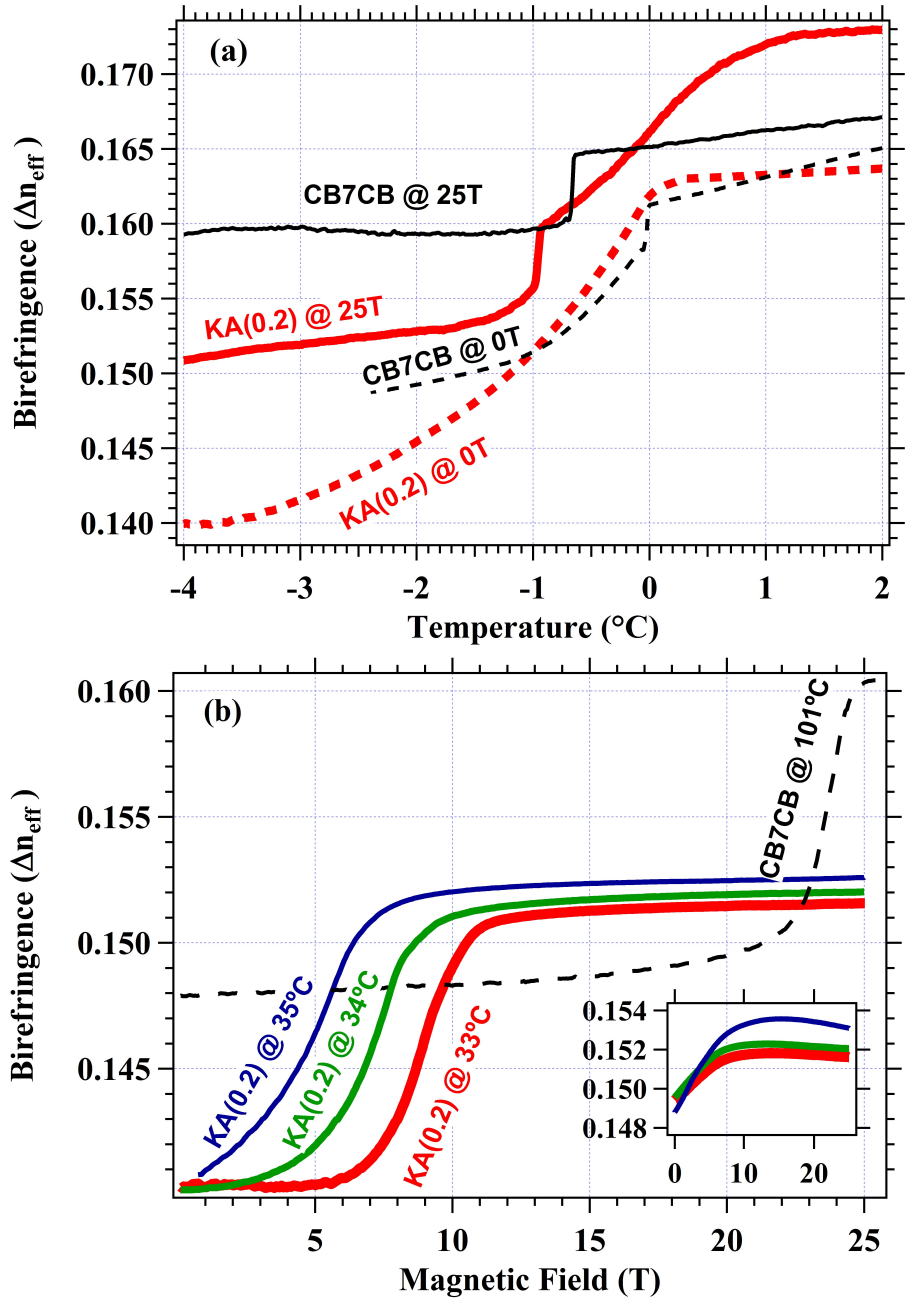

*Figure 3: Effective birefringence measurements for both materials. (a): Temperature dependence at 0 and 25T on cooling at 1°C/min rate; (b): Magnetic field dependences (inset shows the results in decreasing fields).*

The depressions of the N-$N_{tb}$ transition temperatures by the 25T field are due to the positive diamagnetic anisotropy, $\Delta\chi_N$. High field aligns the director uniformly and hence stabilizes the N phase to temperatures lower than the zero-field transition temperature. The shift of the transition temperature, $\Delta T$, induced by the magnetic induction $\vec{B}$, can be estimated using the modified Kirkwood-Helfrich equation [24]: $\frac{\Delta T}{T_o} \cdot \frac{L_M \rho}{M_m} \approx \frac{\Delta\chi_N}{2\mu_o}|\vec{B}|^2$. Here $T_o$ is the zero field transition temperature in Kelvin; $M_m$ is the molar mass; $L_m$ is the latent heat of the transition; $\rho \sim 10^3 \text{kg/m}^3$ is the mass density and $\mu_o = 4\pi \cdot 10^{-7} Vs/(Am)$. For CB7CB $\Delta\chi_N$ should be close to that of pentyl-cyanobiphenyl (5CB), which is $\sim 10^{-6}$ in SI [25]. Using an estimate of $L_m \sim 100$ J/mol from heat capacity measurements [9], we arrive at $\Delta T \sim 1°C$, which is close to the measured 0.7°C shift. The value of $\Delta\chi_N$ is not presently available for KA(0.2), although it should not be too different than that for CB7CB.

Measurements of $\Delta n$ in both phases allow us to estimate the helicone angle $\theta_o$ from the relation $\Delta n_{TB} \approx \Delta n_N \left(1 - \frac{3}{2}\theta_o^2\right)$ [16], where $\Delta n_{TB}$ and $\Delta n_N$ can be determined from $\Delta n_{eff}$ measured in 25T that eliminated the optical stripes. By using the $\Delta n_N$ values (0.173 for KA(0.2) and 0.167 for CB7CB) measured before the pretransitional decrease, and $\Delta n_{tb}$ measured 0.5°C below the N-$N_{tb}$ transition (0.153 for KA(0.2) and 0.159 for CB7CB), we find $\theta_0 \approx 16°$ for KA(0.2) and $\theta_0 \approx 10°$ for CB7CB.

The textures of stripes and focal conic domains are ubiquitous in systems such as smectics and cholesterics, formed by one-dimensional stacking of flexible equidistant layers [1][26]. They indicate the tendency of layers to tilt in response to a mechanical stress [26], such as the temperature-induced variation of the period. The underlying physics is captured well by the so-called Helfrich-Hurault (HH) buckling instability [1], which usually is described for the geometry in which the layers are parallel to the bounding plates of the cell, however the general mechanism of the HH buckling remains valid even when the layers are perpendicular to the bounding plates. In such "bookshelf" geometry, the tilt of smectic layers typically occurs across the cell (vertical chevron) [28], but it is also observed in the plane of the cell [27][28][29]. The latter "horizontal chevron" texture is similar to the optical stripes observed in $N_{tb}$ that, thanks to the nanoscale twist-bend deformations, also has a pseudo layer structure. The optical stripes shown in Fig. 4(a) and (b) are parallel to the average optical axis, and the period of undulating stripe patterns proportional to the cell thickness [6], just as established for a smectic C in the bookshelf geometry. [26] We use the HH-buckling framework to analyze the magnetic field-induced elimination of the optical stripes and to estimate elastic properties of $N_{tb}$.

We assume that the displacement of the undulating pseudo layers of the $N_{tb}$ phase has the form $u = u_0 \sin(q_z z) \cdot \cos(q_x x)$, where the z-axis is parallel to the stripes (helical axis of a uniform state) and the x-axis is directed perpendicularly to the stripes, Figure 4(c).

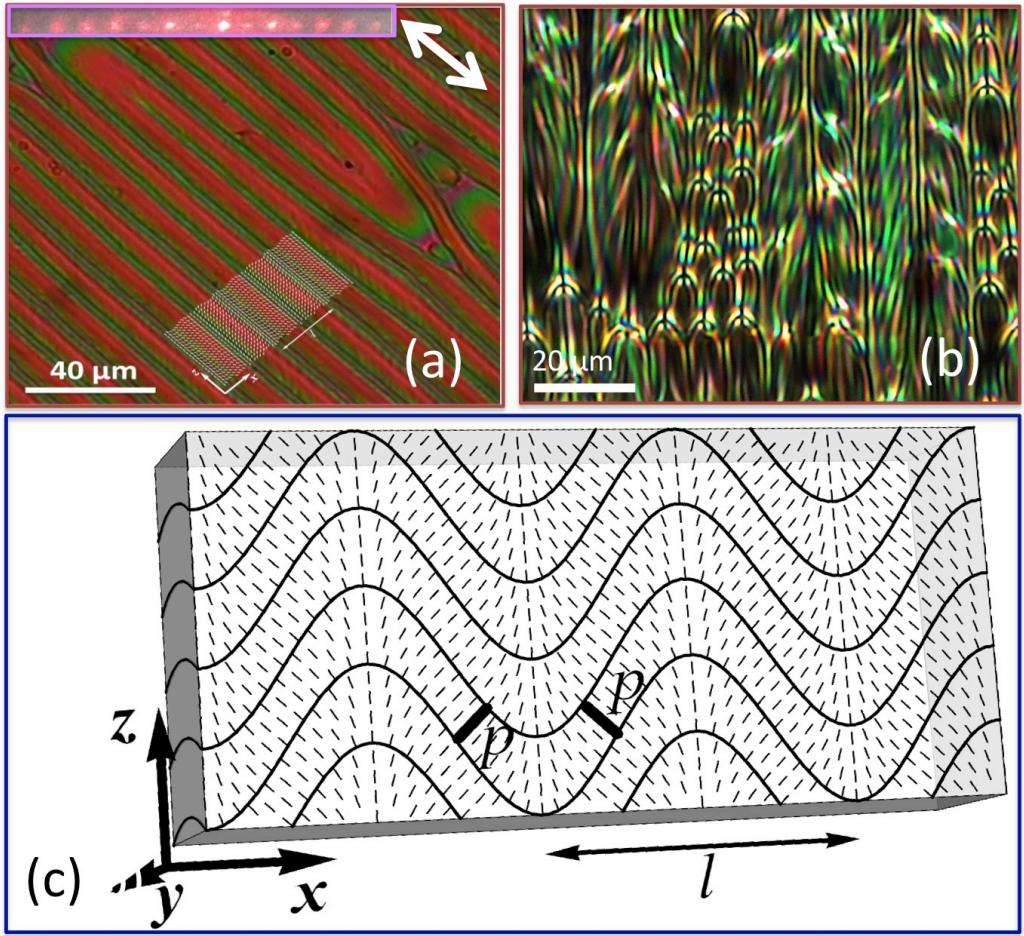

*Figure 4: Illustration and model of the optical stripes. (a): Optical texture of a 10μm thick film of KA(0.2) with planar anchoring between crossed polarizers parallel to the edges of the picture. White arrow indicates the rubbing direction. Inset on top shows the diffraction pattern, and the dotted white lines indicate the modulation of the helical axis. (b): Optical texture of a 10μm thick film of CB7CB; (c) Illustration of the undulation of the pseudo layers determined by the pitch of the conical helix of pitch p. Due to the undulation the direction of the heliconical axes (short lines) are periodically changing in the x direction by a period l.*

The "coarse-grain" model of $N_{tb}$ with weakly distorted pseudo-layers is described by the energy density functional

$$F_{cg} = \frac{1}{2}B\left[\frac{\partial u}{\partial z} - \frac{1}{2}\left(\frac{\partial u}{\partial x}\right)^2\right]^2 + \frac{1}{2}K\left(\frac{\partial^2 u}{\partial x^2}\right)^2 - \frac{1}{2}\frac{\Delta\chi_{TB}|\vec{B}|^2}{\mu_o} \quad (1)$$

Here $B$ is the layer compression modulus, $K$ is the effective splay constant for the helicoidal axis, $\Delta\chi_{TB} = \Delta\chi_N(3\cos^2\theta_0 - 1)/2$ is the diamagnetic susceptibility anisotropy of $N_{tb}$. The HH relation describing the geometry of undulation is $q_x^2 = q_z/\lambda$, where $q_x = 2\pi/l$, $l$ is the observed period of optical stripes, $\lambda = \sqrt{K/B}$ is the elastic extrapolation length, and $q_z$ is set by a less known anchoring conditions; its value does not enter directly into the numerical calculations below. The magnitude of the magnetic induction $|\vec{B}_c|$ that restores the flat configuration of layers can be determined by equating the magnetic and elastic energies in Eq.(1) [1]: $\Delta\chi_{TB}|\vec{B}_c|^2/\mu_o = 2q_x^2 K = 8\pi^2 K/l^2$. The experimental values of $|\vec{B}_c|$ for KA(0.2) are 4T, 6.5T, and 8T at $T_{TB} - T = 2.4°$, 3.4°, and 4.4°C, respectively and 23T at $T_{TB} - T = 2.4°C$ for CB7CB [see Figure 2(b)]. Since $\theta_o$ is small $\Delta\chi_{TB} \sim \Delta\chi_N \sim 10^{-6}$ for both materials, and using the observed $l \sim 10$ μm for KA(0.2) and $l \sim 2\mu m$ for CB7CB, we find that $K$ increases from 20 pN to about 80 pN, as the temperature decreases from $T_{TB} - T = 2.4°$ to 4.4°C for KA(0.2), and of $K \sim 25$ pN in CB7CB at $T_{TB} - T = 2.4°C$. The experimental values $K \approx 20 - 80$ pN for the $N_{tb}$ phase of KA(0.2) are higher than the splay elastic constant $K_1 \approx 15$ pN measured in the N phase of this material near the N-$N_{tb}$ transition point [14], which is an expected result. It would be of interest to measure the compressibility modulus $B$ in Eq.(1) directly, for example, by causing undulations of pseudo-layers in dilated homeotropic samples, as described for smectic and cholesteric liquid crystals. [30,31] One would expect $B$ in the $N_{TB}$ phase to be smaller than its counterpart in the thermotropic smectic liquid crystals, as the change in pseudo-layers spacing implies essentially molecular realignments within the oblique helicoidal director configuration rather than a change in the molecular density.

There are two features of the $N_{tb}$ phase that make the coarse-grained models such as Eq.(1) an attractive option for the description of deformed states of the $N_{tb}$ phase, such as the striped textures or field-induced Frederiks transition involving splay and saddle-splay deformations, as discussed in Ref. [15]. First, the $N_{tb}$ pitch ($p_o\sim10nm$) [16] [20] is very small, not much larger than the molecular length (~3nm). Second, $N_{tb}$ is expected to feature coexisting left- and right-hand twisted domains [3] which implies a presence of walls separating them [15] and thus spatial heterogeneity at various scales, from about 30 nm [15] to microns [18], depending on the sample and its prehistory.

In summary, we have demonstrated that a high magnetic field shifts the N-$N_{tb}$ transition temperature and suppresses the micron-scale ("optical") stripe patterns of the $N_{tb}$ samples. We explain the optical stripes as undulations of the Helfrich-Hurault type inherent to other modulated liquid crystal phases, such as smectics and short-pitch cholesterics. The suppression of stripes by the magnetic field is explained by the coarse-grained model that treats the $N_{tb}$ phase as a system of pseudo-layers and allows us to estimate the splay elastic constant of the $N_{tb}$ hecicoidal axis that is a few times larger than the splay elastic constant of the corresponding nematic phase.

## Acknowledgement

This work was supported by the NSF under grants DMR 0964765, 1104805, 1121288 and 1307674, as well as the DOE grant DE-FG02-06ER 46331. Invaluable assistance was provided by W. Aldhizer and S.W. McGill. Work performed at NHMFL supported by NSF cooperative agreement DMR-0084173, the State of Florida and the U.S. Department of Energy.